\documentclass[preprin,showpacs,preprintnumbers,eqsecnum,amsmath,amssymb]{revtex4}
\usepackage{graphics,epsfig}
\usepackage{graphicx}
\usepackage{dcolumn}
\usepackage{bm}

\begin{document}
\baselineskip=0.8 cm
\title{{\bf Strong gravitational lensing in a noncommutative black-hole spacetime}}

\author{Chikun Ding }
\email{dingchikun@163.com} \affiliation{Department of Physics and Information Engineering, \\
Hunan Institute of Humanities Science and Technology,\\ Loudi, Hunan
417000, P. R. China}
\author{ Shuai Kang}
 \affiliation{Department of Physics and Information Engineering, \\
Hunan Institute of Humanities Science and Technology,\\ Loudi, Hunan
417000, P. R. China}
\author{ Chang-Yong Chen}
 \affiliation{Department of Physics and Information Engineering, \\
Hunan Institute of Humanities Science and Technology,\\ Loudi, Hunan
417000, P. R. China}
\author{Songbai Chen }
\email{csb3752@163.com}
 \affiliation{Institute of Physics and
Department of Physics,
Hunan Normal University,  Changsha, Hunan 410081, P. R. China \\
Key Laboratory of Low Dimensional Quantum Structures and Quantum
Control (Hunan Normal University), Ministry of Education, P. R.
China.}

\author{Jiliang Jing }
\email{jljing@hunnu.edu.cn}
 \affiliation{Institute of Physics and
Department of Physics,
Hunan Normal University,  Changsha, Hunan 410081, P. R. China \\
Key Laboratory of Low Dimensional Quantum Structures and Quantum
Control (Hunan Normal University), Ministry of Education, P. R.
China.}

\vspace*{0.2cm}
\begin{abstract}
\baselineskip=0.6 cm
\begin{center}
{\bf Abstract}
\end{center}

Noncommutative geometry may be a starting point to a quantum gravity.
We study the influence of the spacetime noncommutative parameter on
the strong field gravitational lensing in the noncommutative
Schwarzschild black-hole spacetime and obtain the angular position
and magnification of the relativistic images. Supposing that the
gravitational field of the supermassive central object of the galaxy
described by this metric, we estimate the numerical values of the
coefficients and observables for strong gravitational lensing.
Comparing to the Reissner-Norstr\"{om} black hole, we find that the
influences of the spacetime noncommutative parameter is similar to those of the
charge, just these influences are much smaller. This may offer a way
to distinguish a noncommutative black hole from a
Reissner-Norstr\"{om} black hole, and may probe the spacetime
noncommutative constant $\vartheta$ \cite{foot} by the astronomical
instruments in the future.
\end{abstract}

 \maketitle
\newpage
\vspace*{0.2cm}
\section{Introduction}

The theoretical discovery of radiating black holes disclosed the
first window on the mysteries of quantum gravity. Though after thirty years
of intensive research, the full quantum gravity is still unknown.
 However there are two candidates for quantum gravity,
which are the string theory and the loop quantum gravity. By the
string/black hole correspondence principle \cite{suss}, stringy
effects cannot be neglected in the late stage of a black hole. In
the string theory, coordinates of the target spacetime become {\it
noncommutating} operators on a $D$-brane as \cite{seigerg}
\begin{equation}[\hat{x}^\mu,\hat{x}^\nu]=i\vartheta^{\mu\nu},\end{equation} where
$\vartheta^{\mu\nu}$  is a real, anti-symmetric and constant tensor
which determines the fundamental cell discretization of spacetime
much in the same way as the Planck constant $\hbar$ discretizes the
phase space, $[\hat{x}_i,\hat{p}_j]=i\hbar\delta_{ij}$. Motivated by
string theory arguments, noncommutative spacetime has been
reconsidered again and is believed to afford a starting point to quantum gravity.

Noncommutative spacetime is not a new conception, and coordinate
noncommutativity also appears in another fields, such as in quantum
Hall effect \cite{bell}, cosmology \cite{marcoll}, the model of a
very slowly moving charged particle on a constant magnetic field
\cite{kim}, the Chern-Simon's theory \cite{deser}, and so on.
 The idea of
noncommutative spacetime dates back to Snyder \cite{snyder} who used
the noncommutative structure of spacetime to introduce a small
length scale cut-off in field theory without breaking Lorentz
invariance and Yang \cite{yang} who extended Snyder's work to
quantize spacetime in 1947 before the renormalization theory.
Noncommutative geometry \cite{connes} is a branch of mathematics
that has many applications in physics, a good review of the
noncommutative spacetime is in \cite{connes2, akofor}.

The fundamental notion of the noncommutative geometry is that the
picture of spacetime as a manifold of points breaks down at distance
scales of the order of the Planck length: Spacetime events cannot be
localized with an accuracy given by Planck length \cite{akofor} as
well as particles do in the quantum phase space. So that the points on
the classical commutative manifold should then be replaced by states
on a noncommutative algebra and the point-like object is replaced by
a smeared object \cite{smail} to cure the singularity problems at
the terminal stage of black hole evaporation \cite{nico}.

The approach to noncommutative quantum field theory follows two
paths: one is based on the Weyl-Wigner- Moyal *-product and the
other on coordinate coherent state formalism \cite{smail}. In a
recent paper, following the coherent state approach, it has been
shown that Lorentz invariance and
unitary, which are controversial questions raised in the *-product approach \cite{morita}, can be
achieved by assuming\begin{equation}\vartheta^{\mu\nu}=\vartheta \;
\text{diag}(\epsilon_1,\ldots, \epsilon_{D/2}),\end{equation} where
$\vartheta$ \cite{foot} is a constant which has the dimension of $length^2$, $D$ is the dimension of
spacetime \cite{smail2} and, there isn't any UV/IR mixing.
Inspire by these results, various black hole solutions of noncommutative spacetime have been
found \cite{Nicolini:2008aj}; thermodynamic properties of the
noncommutative black hole were studied in \cite{nozari}; the
evaporation of the noncommutative black hole was studied in
\cite{comp}; quantized entropy was studied in \cite{wei}, and so on.

It is interesting that  the noncommutative spacetime coordinates introduce a
new fundamental natural length scale $\sqrt{\vartheta}$.
  In this paper, we plan to study the influence of this
constant on strong gravitational lensing.

The earlier studies of gravitational lensing have been developed in
the weak field approximation \cite{Schneider}-\cite{RDB}. It is
enough for us to investigate the properties of gravitational lensing
by ordinary stars and galaxies. However, when the lens is a black
hole, a strong field treatment of gravitational lensing
\cite{Darwin,Vir,Vir1,Vir2,Vir3,Fritt} is need instead. Virbhadra and
Ellis \cite{Vir1} find that near the line connecting the source and
the lens, an observer would detect two infinite sets of faint
relativistic images on each side of the black hole. These relativistic images could
provide a profound verification of alternative theories of gravity.
Thus, the study of the strong gravitational lensing becomes
appealing recent years. On the basis of the Virbhadra-Ellis lens equation \cite{Vir2,Vir3},
Bozza \cite{Bozza2} extended the analytical method of lensing for a
general class of static and spherically symmetric spacetimes and
showed that the logarithmic divergence of the deflection angle at
photon sphere is a common feature. Then Bhadra
\textit{et al} \cite{Bhad1}\cite{Sarkar} have considered the
Gibbons-Maeda-Garfinkle-Horowitz-Strominger black hole lensing. Eiroa \textit{et al} \cite{Eirc1} have studied the
Reissner-Nordstr\"{o}m black hole lensing. Konoplya \cite{Konoplya1}
has studied the corrections to the deflection angle and time delay
of black hole lensing immersed in a uniform magnetic
field. Majumdar \cite{Muk} has investigated the dilaton-de Sitter black hole lensing. Perlick
\cite{Per} has obtained an exact lens equation and used it to study
 Barriola-Vilenkin monopole black hole lensing. S. Chen studied the K-S black hole lensing \cite{chen}.
 Bin-Nun
\cite{bin} studied the strong gravitational lensing by Sgr A*, and
so on.

The plan of our paper is organized as follows. In Sec. II we adopt to
Bozza's method and obtain the deflection angles for light rays
propagating in the noncommutative Schwarzschild black hole
spacetime. In Sec. III we suppose that the gravitational field of the
supermassive black hole at the centre of our galaxy can be described
by this metric and then obtain the numerical results for the
observational gravitational lensing parameters defined in Sec. II.
Then, we make a comparison between the properties of gravitational
lensing in the noncommutative Schwarzschild and
Reissner-Norstr\"{om} metrics. In Sec. IV, we present a summary.

\section{Deflection angle in the noncommutative Schwarzschild black hole spacetime}

The line element of the noncommutative Schwarzschild black hole
reads \cite{nico}
\begin{eqnarray}
ds^2 = -f(r)\,dt^2 + \frac{dr^2}{f(r)} + r^2 (d\theta^2
+\sin^2\theta d\phi^2)\,, \label{metric}
\end{eqnarray}
and
\begin{eqnarray}
\label{sol1} f(r)=1- \frac{4M}{r\sqrt{\pi}}\, \gamma(3/2 \ ,
r^2/4\vartheta\,)\,,
\end{eqnarray}
where $\gamma\left(3/2 \ , r^2/4\vartheta\, \right)$ is the lower
incomplete Gamma function:
\begin{equation}
\gamma\left(3/2\ , r^2/4\vartheta\, \right)\equiv
\int_0^{r^2/4\vartheta} dt\, t^{1/2} e^{-t},
\end{equation}
$\vartheta$ is a spacetime noncommutative parameter \cite{foot}. The
commutative Schwarzschild metric is obtained from (\ref{metric}) in
the limit $r/\sqrt{\vartheta}\to\infty $. And Eq.(\ref{metric})
leads to the mass distribution $m\left(\, r\,\right)= 2M
\,\gamma\left(3/2\ , r^2/4\vartheta\, \right)/\sqrt\pi $, where $M$
is the total mass of the source.  When $M>1.9\sqrt{\vartheta}$, the
event horizons are given by
\begin{eqnarray}
r_\pm=\frac{4M}{\sqrt{\pi}}\,\gamma\left(3/2\ , r^2_\pm/4\vartheta\,
\right),
\end{eqnarray}which behaviors as that of Reissner-Norstr\"{om} black hole.
The line element (\ref{metric}) describes the geometry of a
noncommutative black hole and should give us useful insights about
possible spacetime noncommutative effects on strong gravitational
lensing.

As in \cite{Vir2,Vir3,Bozza2}, we set $2M=1$ and rewrite the metric
(\ref{metric}) as
\begin{eqnarray}
ds^2=-A(r)dt^2+B(r)dr^2+C(r)\Big(d\theta^2+\sin^2\theta
d\phi^2\Big),\label{grm}
\end{eqnarray}
with
\begin{eqnarray}
A(r)=f(r), \;\;\;\;B(r)&=&1/f(r),\;\;\;\; C(r)=r^2.
\end{eqnarray}
The deflection angle for the photon coming from infinite can be
expressed as
\begin{eqnarray}
\alpha(r_0)=I(r_0)-\pi,
\end{eqnarray}
where $r_0$ is the closest approach distance and $I(r_0)$ is
\cite{Vir2,Vir3}
\begin{eqnarray}
I(r_0)=2\int^{\infty}_{r_0}\frac{\sqrt{B(r)}dr}{\sqrt{C(r)}
\sqrt{\frac{C(r)A(r_0)}{C(r_0)A(r)}-1}}.\label{int1}
\end{eqnarray}
It is easy to obtain that as parameter $r_0$ decrease the deflection
angle increase. At certain a point, the deflection angle will become
$2\pi$, it means that the light ray will make a complete loop around
the compact object before reaching the observer. When $r_0$ is equal
to the radius of the photon sphere, the deflection angle diverges
and the photon is captured.

The photon sphere equation is given by \cite{Vir2,Vir3}
\begin{eqnarray}
\frac{C'(r)}{C(r)}=\frac{A'(r)}{A(r)},\label{root}
\end{eqnarray}
which admits at least one positive solution and then the largest
real root of Eq.(\ref{root}) is defined  as the radius of the photon
sphere. To the noncommutative Schwarzschild black hole metric
(\ref{metric}), the radius of the photon sphere can be given
implicitly by
\begin{eqnarray}\label{rpseq}
r_{ps}=\frac{3}{2}-\left[\frac{r_{ps}^3}{4\vartheta\sqrt{\pi\vartheta}}
e^{-\frac{r_{ps}^2}{4\vartheta}}+\frac{3}{\sqrt{\pi}}\Gamma(\frac{3}{2},
\frac{r_{ps}^2}{4\vartheta})\right],
\end{eqnarray}
which is an implicit function $f(r_{ps},\vartheta)=0$. It cannot be
expressed as explicit function $r_{ps}=g(\vartheta)$, so we list
some values of the photon sphere radius in the following table, and
describe them in the Fig. 1.
\begin{table}[h]\label{tabel0}
\caption{Numerical values for the radius of the photon sphere in the
noncommutative Schwarzschild black hole spacetime with different
$\sqrt{\vartheta}$. }
\begin{center}
\begin{tabular}{|c|c|c|c|c|c|c|c|}
\hline \hline $\sqrt{\vartheta}$ &0.260&0.254 & 0.248 &0.242&
0.236&0.230&0.224 \\
\hline
 $r_{ps}$& 1.49151& 1.49405&1.49593&1.49721& 1.49824& 1.49890&1.49934\\
 \hline $\sqrt{\vartheta}$ &0.218&0.212&0.206 & 0.200 &0.194&
0.188&0.182 \\
\hline
 $r_{ps}$& 1.49962& 1.49979&1.49989&1.49995& 1.49998& 1.49999&1.50000\\
\hline\hline
\end{tabular}
\end{center}
\end{table}
\begin{figure}[ht]
\begin{center}
\includegraphics[width=9.0cm]{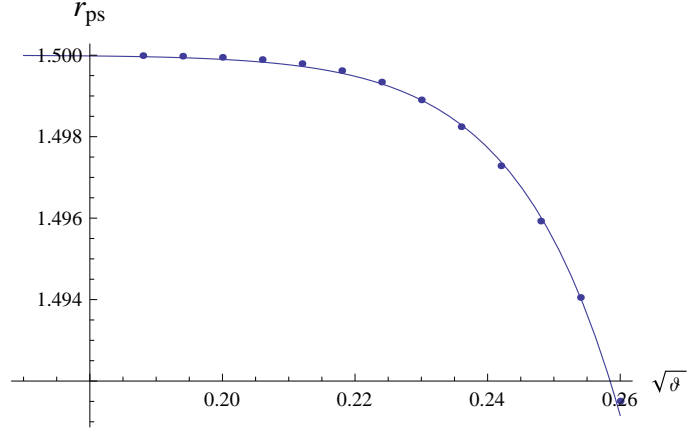}\;\;\;\;\;
\caption{The figure is for the radius of the photon sphere in the
noncommutative Schwarzschild black hole spacetime with different
$\sqrt{\vartheta}$. The dots are the exactly values described by
Tab. I, the line is described by the expression
$r_{ps}=1.5-7.8\times10^7\sqrt{\vartheta}^{17}$.}
 \end{center}
 \label{fp1}
 \end{figure}
From the Tab. I, when $\sqrt{\vartheta}\rightarrow0$, it can
recovers that in the commutative Schwarzschild black hole spacetime
which $r_{ps}=1.5$. Fig. 1 shows that the relation between the
photon sphere radius and the spacetime noncommutative parameter
$\vartheta$ is very coincident to the function
\begin{eqnarray}\label{rpsline}
r_{ps}=1.5-7.8\times10^7\sqrt{\vartheta}^{17},\;\;\;\sqrt{\vartheta}\in(0,\frac{1}{3.8}).
\end{eqnarray}
 It is easy to see that this relation is quite different from that in
the Reissner-Norstr\"{om} black hole spacetime
$r_{ps}=(3+\sqrt{9-32q^2})/4$, which implies that there exist some
distinct effects of the noncommutative parameter $\vartheta$ on
gravitational lensing in the strong field limit.

Following the method developed by Bozza \cite{Bozza2,chen}, we
define a variable
\begin{eqnarray}
z=1-\frac{r_0}{r},
\end{eqnarray}
and obtain
\begin{eqnarray}
I(r_0)=\int^{1}_{0}R(z,r_0)f(z,r_0)dz,\label{in1}
\end{eqnarray}
where
\begin{eqnarray}
R(z,r_0)&=&\frac{2r_0\sqrt{A(r)B(r)C(r_0)}}{C(r)(1-z)^2}=2,
\end{eqnarray}
\begin{eqnarray}
f(z,r_0)&=&\frac{1}{\sqrt{A(r_0)-A(r)C(r_0)/C(r)}}.
\end{eqnarray}
The function $R(z, r_0)$ is regular for all values of $z$ and $r_0$.
However, $f(z, r_0)$ diverges as $z$ tends to zero. Thus, we split
the integral (\ref{in1}) into two parts
\begin{eqnarray}
I_D(r_0)&=&\int^{1}_{0}R(0,r_{ps})f_0(z,r_0)dz, \nonumber\\
I_R(r_0)&=&\int^{1}_{0}[R(z,r_0)f(z,r_0)-R(0,r_{ps})f_0(z,r_0)]dz
\label{intbr},
\end{eqnarray}
where $I_D(r_0)$ and $I_R(r_0)$ denote the divergent and regular
parts in the integral (\ref{in1}), respectively. To find the order
of divergence of the integrand, we expand the argument of the square
root in $f(z,r_0)$ to the second order in $z$ and obtain the
function $f_0(z,r_0)$:
\begin{eqnarray}
f_0(z,r_0)=\frac{1}{\sqrt{p(r_0)z+q(r_0)z^2}},
\end{eqnarray}
where
\begin{eqnarray}
p(r_0)&=&2-\frac{3}{r_0}+\frac{6}{\sqrt{\pi}r_0}\Gamma(\frac{3}{2},\frac{r_0^2}{4\vartheta}
)+\frac{r_0^2}{2\vartheta\sqrt{\pi\vartheta}}e^{-\frac{r_0^2}{4\vartheta}},\;
 \nonumber\\
q(r_0)&=&\frac{3}{r_0}-1-\frac{6}{\sqrt{\pi}r_0}\Gamma(\frac{3}{2},\frac{r_0^2}{4\vartheta}
)-\frac{r_0^2}{4\vartheta\sqrt{\pi\vartheta}}e^{-\frac{r_0^2}{4\vartheta}}
\Big(2+\frac{r_0^2}{2\vartheta}\Big).
\end{eqnarray}
When $r_0$ is equal to the radius of photon sphere $r_{ps}$, the
coefficient $p(r_0)$ vanishes and the leading term of the divergence
in $f_0(z,r_0)$ is $z^{-1}$, thus the integral (\ref{in1}) diverges
logarithmically. Close to the divergence, Bozza \cite{Bozza2} found
that the deflection angle can be expanded in the form
\begin{eqnarray}
\alpha(\theta)=-\bar{a}\log{\bigg(\frac{\theta
D_{OL}}{u_{ps}}-1\bigg)}+\bar{b}+O(u-u_{ps}),
\end{eqnarray}
where
\begin{eqnarray}
&\bar{a}&=\frac{R(0,r_{ps})}{2\sqrt{q(r_{ps})}}=\Big[1-\frac{r_{ps}^4}{8\vartheta^2\sqrt{\pi\vartheta}
}e^{-\frac{r_{ps}^2}{4\vartheta}}\Big]^{-\frac{1}{2}}, \nonumber\\
&\bar{b}&=
-\pi+b_R+\bar{a}\log{\frac{4q^2(r_{ps})\big[2A(r_{ps})-r_{ps}^2A''(r_{ps})\big]}{
p^{'2}(r_{ps})u_{ps}r_{ps}\sqrt{A^3(r_{ps})}}}, \nonumber\\
&b_R&=I_R(r_{ps}),\;\;\;\;\;p'(r_{ps})=\frac{dp}{dr_0}\big|_{r_0=r_{ps}},\;\;\;\;\;u_{ps}
=\frac{r_{ps}}{\sqrt{A(r_{ps})}}.
\end{eqnarray}
$D_{OL}$ denotes the distance between the observer and the
gravitational lens, $\bar{a}$ and $\bar{b}$ are so-called the strong
field limit coefficients which depend on the metric functions
evaluated at $r_{ps}$. In general, the coefficient $b_R$ can not be
calculated analytically and, in this case it cannot be evaluated
numerically. Here we expand the integrand in (\ref{intbr}) in powers
of $\sqrt{\vartheta}$ as in \cite{Bozza2}. Because the values of
various low derivative of integrand of $I_R(r_{ps})$ at
$\vartheta\rightarrow0$ is zero, we can get
\begin{eqnarray}
b_R=2\log[6(2-\sqrt{3})]+ O(\sqrt{\vartheta}).
\end{eqnarray}
Then we can obtain the $\bar{a}$, $\bar{b}$ and $u_{ps}$, and
describe them in Fig (2).
\begin{figure}[ht]
\begin{center}
\includegraphics[width=5.0cm]{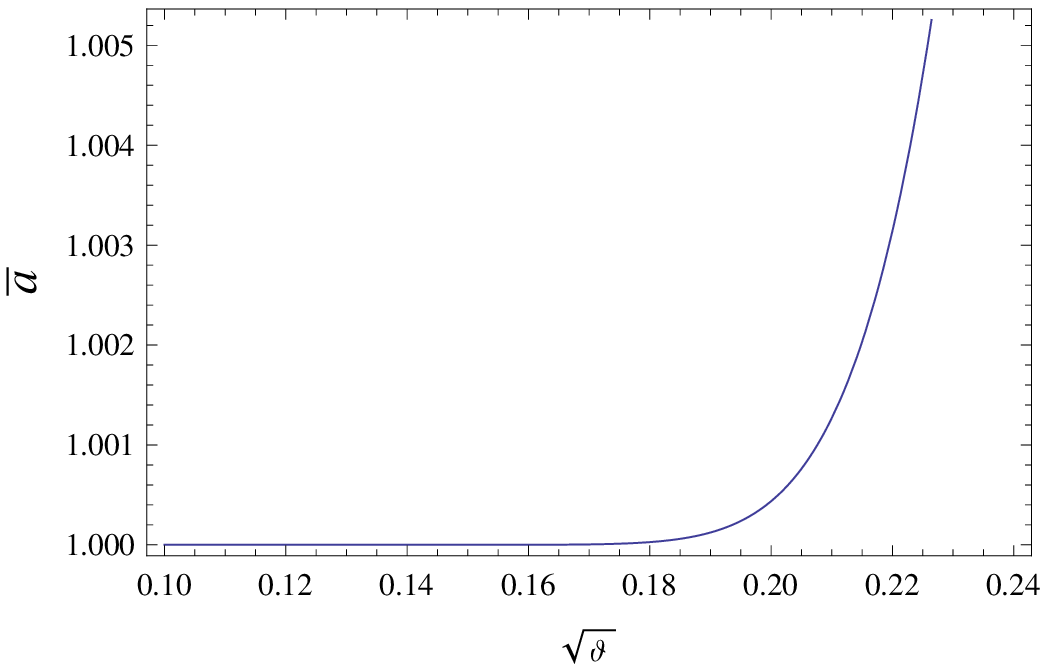}\;\;\;\;\;
 \includegraphics[width=5.0cm]{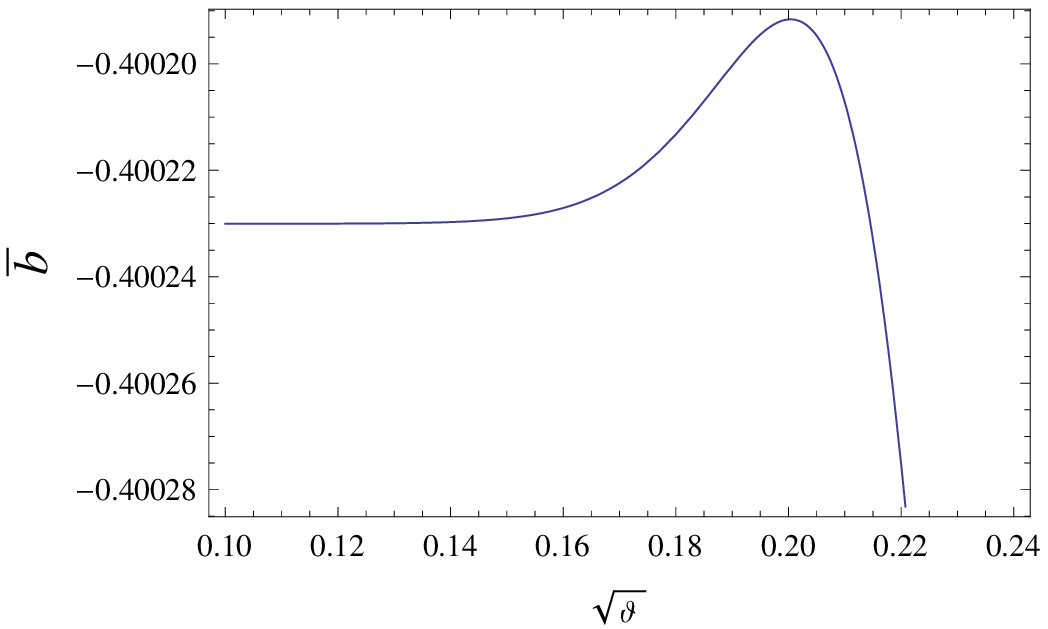}\;\;\;\;\includegraphics[width=5.0cm]{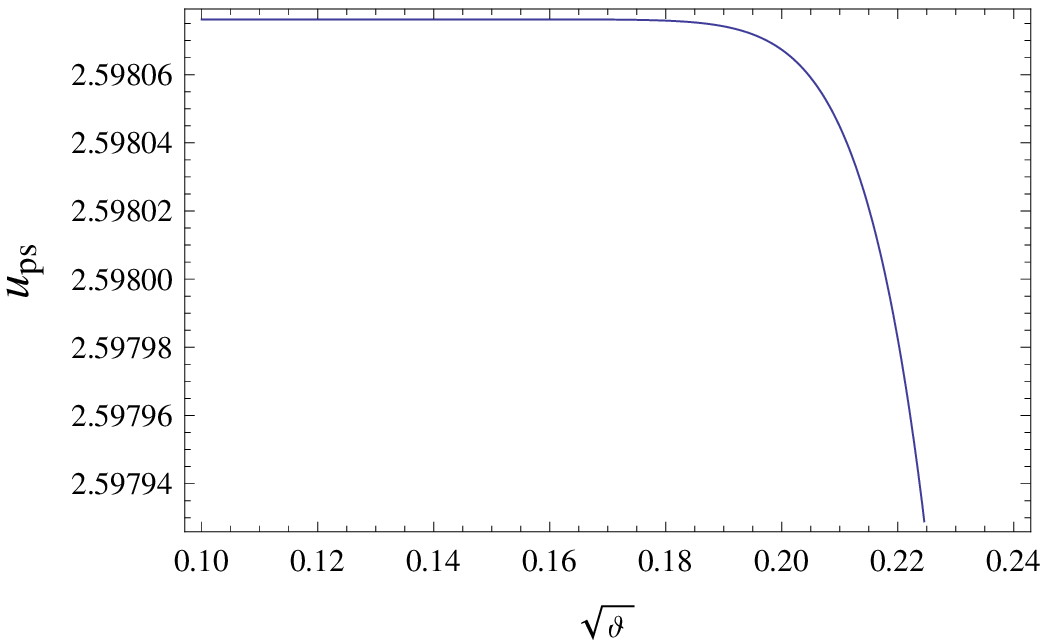}\\
 \includegraphics[width=5.0cm]{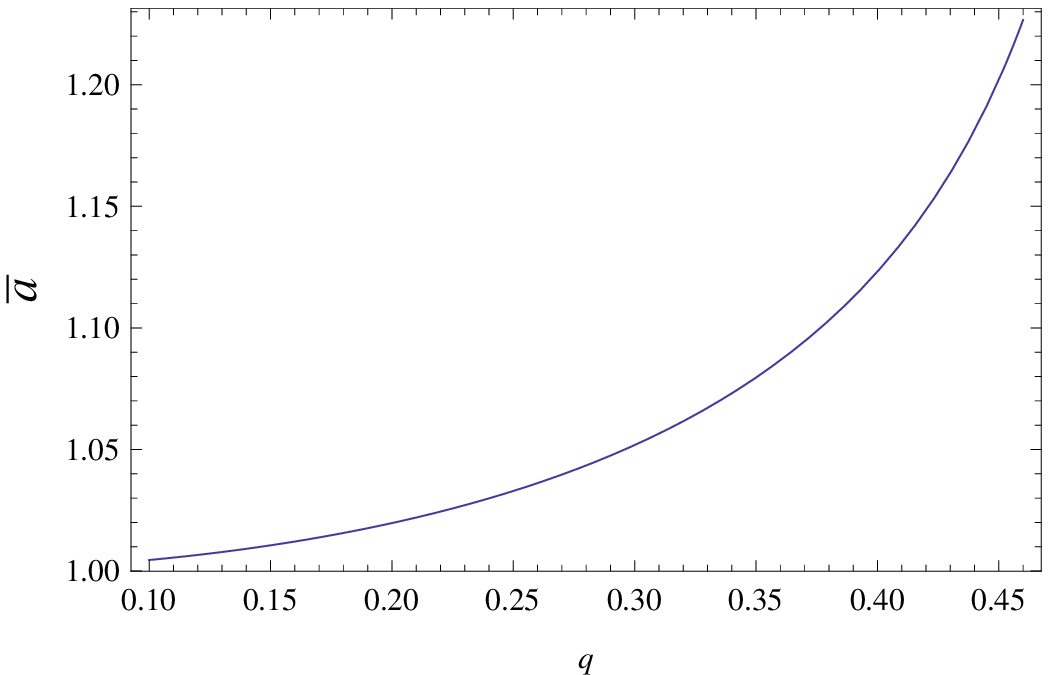}\;\;\;\;\;
 \includegraphics[width=5.0cm]{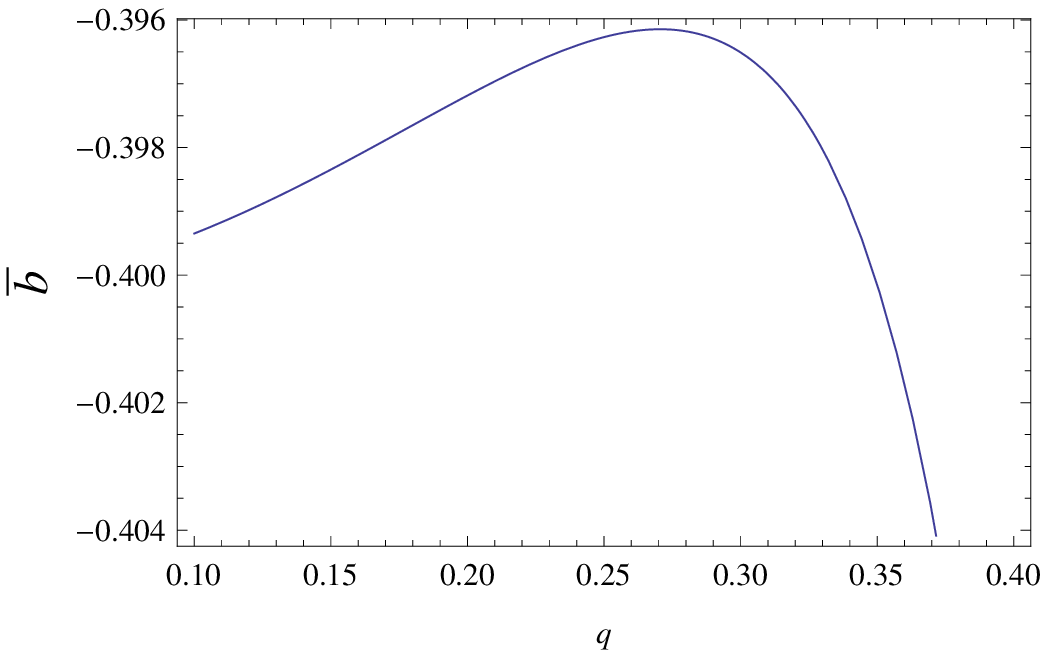}\;\;\;\;\includegraphics[width=5.0cm]{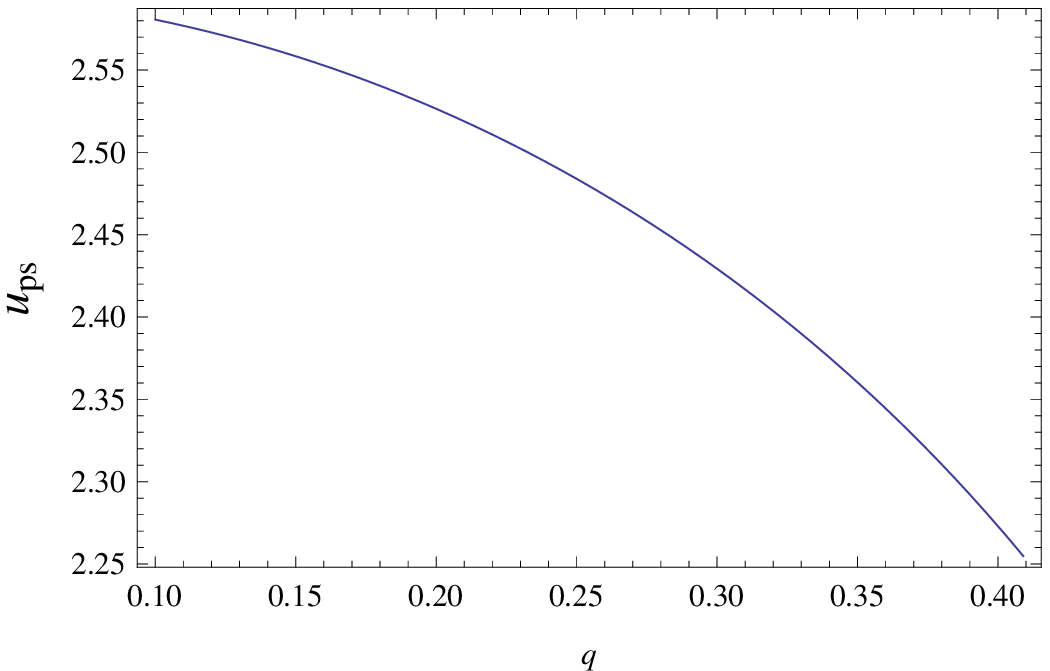}
\caption{Variation of the coefficients of the strong field limit
$\bar{a}$, $\bar{b}$ and the minimum impact parameter $u_{ps}$ with
the spacetime noncommutative parameter $\sqrt{\vartheta}$ in the
noncommutative Schwarzschild black hole spacetime (in the upper row)
and with $q$ in the Reissner-Nordstr\"{o}m black hole spacetime (in
the lower row). The values of the coefficients of
Reissner-Nordstr\"{o}m lensing come from \cite{Bozza2}.}
 \end{center}
 \label{f2}
 \end{figure}
 Figures (2) tell us that with
the increase of $\vartheta$ the coefficient $\bar{a}$ increase, the
$\bar{b}$ slowly increases at first, then decrease quickly when it
arrives at a peak, and the minimum impact parameter $u_{ps}$
decreases, which is similar to that in the Reissner-Norstr\"{om}
black hole metric. However, as shown in Fig. (2), in the
noncommutative Schwarzschild black hole, $\bar{a}$ increases more
slowly, both of $\bar{b}$ and $u_{ps}$ decrease more slowly. In a
word, comparing to the Reissner-Nordstrom black hole, the influences
of the spacetime noncommutative parameter on the strong
gravitational lensing is similar to those of the charge, merely they
are much smaller. On the other side, in principle we can distinguish
a noncommutative Schwarzschild black hole from the
Reissner-Nordstrom black hole and, may be probe the value of the spacetime
noncommutative constant by using strong field gravitational lensing.

Figure (3) shows the deflection angle $\alpha (\theta)$ evaluated at
$u=u_{ps}+0.00326$. It indicates that the presence of $\vartheta$
increases the deflection angle $\alpha (\theta)$ for the light
propagated in the noncommutative Schwarzschild black hole spacetime.
Comparing with those in the commutative one, we could extract the
information about the size of spacetime noncommutative parameter
$\vartheta$ by using strong field gravitational lensing.
\begin{figure}[ht]
\begin{center}
\includegraphics[width=9.0cm]{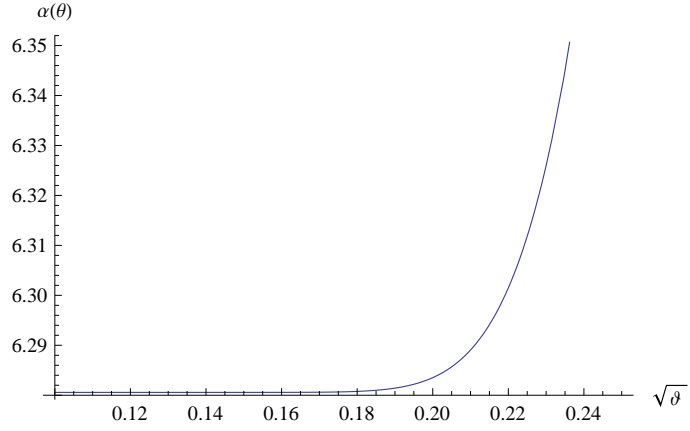}
\caption{ Deflection angles in the noncommutative Schwarzschild
black hole spacetime evaluated at $u=u_{ps}+0.00326$ as functions of
$\sqrt{\vartheta}$.}
 \end{center}
 \label{f1}
 \end{figure}

Considering the source, lens and observer are highly aligned, the
lens equation in strong gravitational lensing can be written as
\cite{Bozza1}
\begin{eqnarray}
\beta=\theta-\frac{D_{LS}}{D_{OS}}\Delta\alpha_{n},
\end{eqnarray}
where $D_{LS}$ is the distance between the lens and the source,
$D_{OS}=D_{LS}+D_{OL}$, $\beta$ is the angular separation between
the source and the lens, $\theta$ is the angular separation between
the image and the lens, $\Delta\alpha_{n}=\alpha-2n\pi$ is the
offset of deflection angle and $n$ is an integer. The position of
the $n$-th relativistic image can be approximated as
\begin{eqnarray}
\theta_n=\theta^0_n+\frac{u_{ps}e_n(\beta-\theta^0_n)D_{OS}}{\bar{a}D_{LS}D_{OL}},
\end{eqnarray}
where
\begin{eqnarray}
e_n=e^{\frac{\bar{b}-2n\pi}{\bar{a}}},
\end{eqnarray}
$\theta^0_n$ are the image positions corresponding to
$\alpha=2n\pi$.  The magnification of $n$-th relativistic image is
given by
\begin{eqnarray}
\mu_n=\frac{u^2_{ps}e_n(1+e_n)D_{OS}}{\bar{a}\beta D_{LS}D^2_{OL}}.
\end{eqnarray}
If $\theta_{\infty}$ represents the asymptotic position of a set of
images in the limit $n\rightarrow \infty$, the minimum impact
parameter $u_{ps}$ can be simply obtained as
\begin{eqnarray}
u_{ps}=D_{OL}\theta_{\infty}.
\end{eqnarray}
In the simplest situation, we consider only that the outermost image
$\theta_1$ is resolved as a single image and all the remaining ones
are packed together at $\theta_{\infty}$. Then the angular
separation between the first image and other ones can be expressed
as
\begin{eqnarray}
s=\theta_1-\theta_{\infty},
\end{eqnarray}
and the ratio of the flux from the first image and those from the
all other images is given by
\begin{eqnarray}
\mathcal{R}=\frac{\mu_1}{\sum^{\infty}_{n=2}\mu_{n}}.
\end{eqnarray}
For highly aligned source, lens and observer geometry, these
observable can be simplified as
\begin{eqnarray}
&s&=\theta_{\infty}e^{\frac{\bar{b}-2\pi}{\bar{a}}},\nonumber\\
&\mathcal{R}&= e^{\frac{2\pi}{\bar{a}}}.
\end{eqnarray}
The strong deflection limit coefficients $\bar{a}$, $\bar{b}$ and
the minimum impact parameter $u_{ps}$ can be obtain through
measuring $s$, $\mathcal{R}$ and $\theta_{\infty}$. Then, comparing
their values with those predicted by the theoretical models, we can
identify the nature of the black hole lens.

\section{Numerical estimation of observational gravitational lensing parameters}

In this section, supposing that the gravitational field of the
supermassive black hole at the galactic center of Milk Way can be
described by the noncommutative Schwarzschild black hole metric, we
estimate the numerical values for the coefficients and observables
of the strong gravitational lensing, and then we study the effect of
the spacetime noncommutative parameter $\vartheta$ on the
gravitational lensing.

The mass of the central object of our Galaxy is estimated to be
$2.8\times 10^6M_{\odot}$ and its distance is around $8.5$ kpc. For
different $\vartheta$, the numerical value of the minimum impact
parameter $u_{ps}$, the angular position of the asymptotic
relativistic images $\theta_{\infty}$, the angular separation $s$
and the relative magnification of the outermost relativistic image
with the other relativistic images $r_{m}$ are listed in the table
(II).
\begin{table}[h]\label{tab1}
\caption{Numerical estimation for main observables and the strong
field limit coefficients for black hole at the center of our galaxy,
which is supposed to be described by the noncommutative
Schwarzschild black hole metric. $R_s$ is Schwarzschild radius.
$r_m=2.5\log{\mathcal{R}}$.}
\begin{center}
\begin{tabular}{|c|c|c|c|c|c|c|}
\hline \hline $\sqrt{\vartheta}$ &$\theta_{\infty}
$($\mu$arcsecs)&\; $s$ ($\mu$arcsecs) \;\; & $r_m$(magnitudes)
&\;\;\;\;$u_{ps}/R_S$\;\;\;\; &
\;\;\;\;\;\;\;\;$\bar{a}$\;\;\;\;\;\;\;\; &\;\;\;\;\;\;\;\;
$\bar{b}$\;\;\;\;\;\;\;\; \\
\hline
 0& 16.870& 0.0211& 6.8219&2.600& 1.000& $-0.4002$ \\
 \hline
0.16&16.8699&0.02109&6.82188&2.59808& 1.00000&$-0.40023$\\
\hline
0.18& 16.8699&0.02110&6.82170&2.59808&1.00003&$-0.40021$ \\
\hline
0.20& 16.8698&0.02116&6.81890&2.59807&1.00044&$-0.40019$ \\
 \hline
0.22 &16.8693&0.02154&6.80052&2.59798&1.00314& $-0.40028$\\
 \hline
0.24&16.8662&0.02304&6.73143&2.59752&1.01344&$-0.40058$\\
 \hline
0.26&16.8550&0.02759&6.54774&2.59579&1.04187&$-0.40019$
 \\
\hline\hline
\end{tabular}
\end{center}
\end{table}\begin{figure}[ht]
\begin{center}
\includegraphics[width=5.0cm]{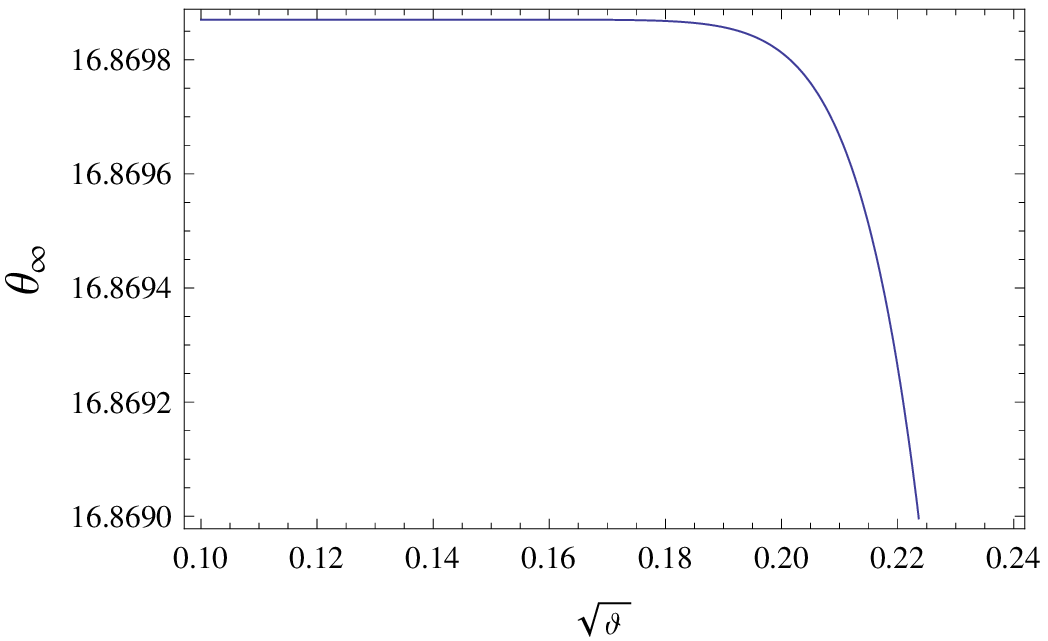}\;\;\;\;\;
 \includegraphics[width=5.0cm]{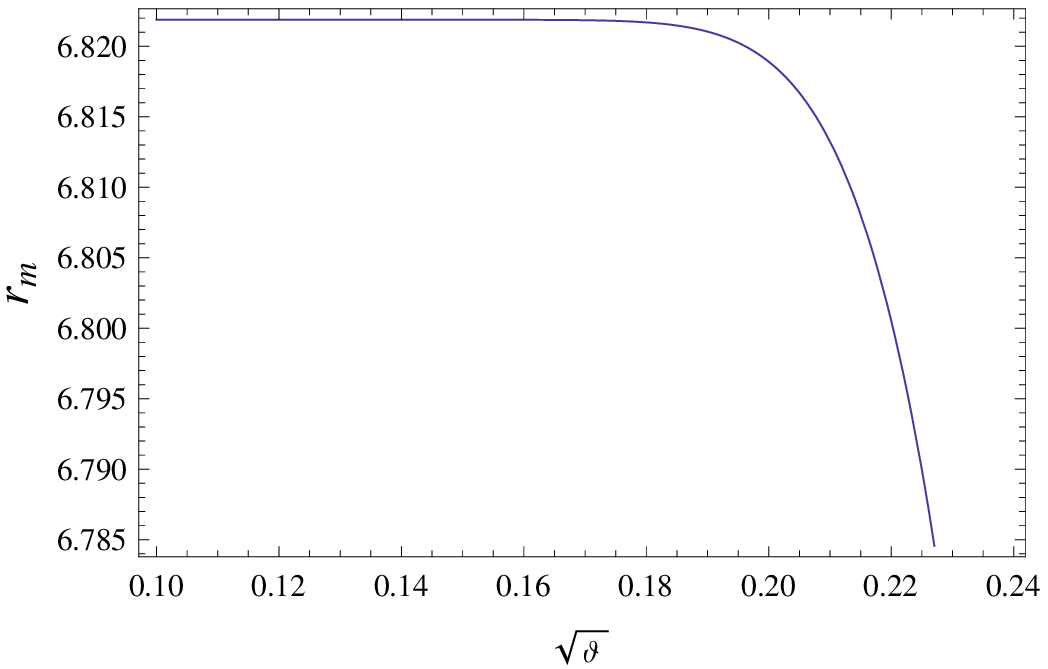}\;\;\;\;\;
 \includegraphics[width=5.0cm]{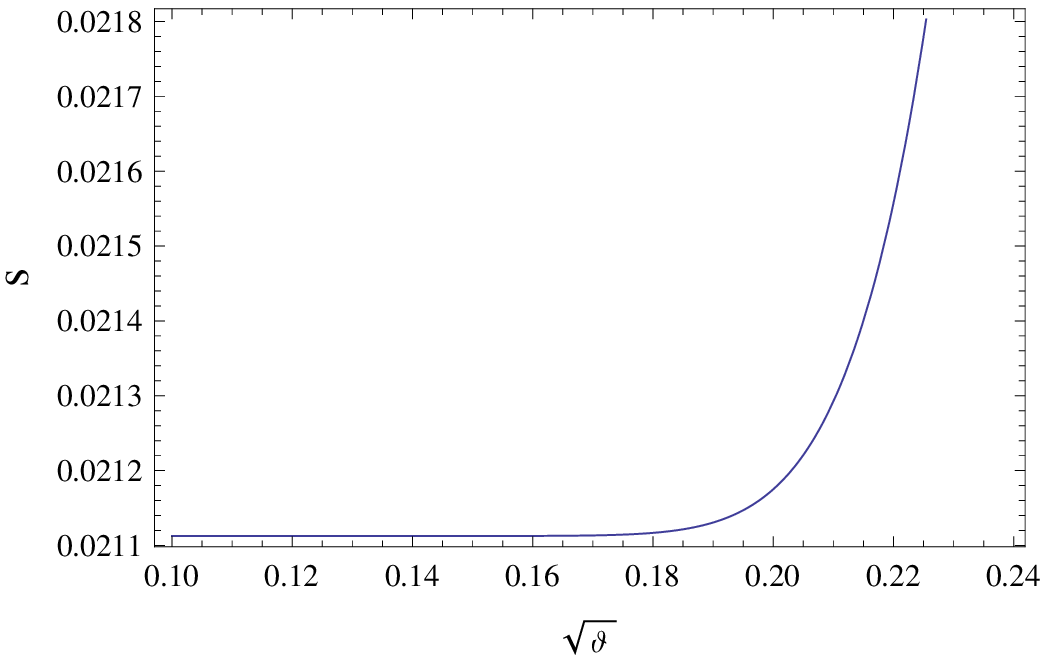}\\
 \includegraphics[width=5.0cm]{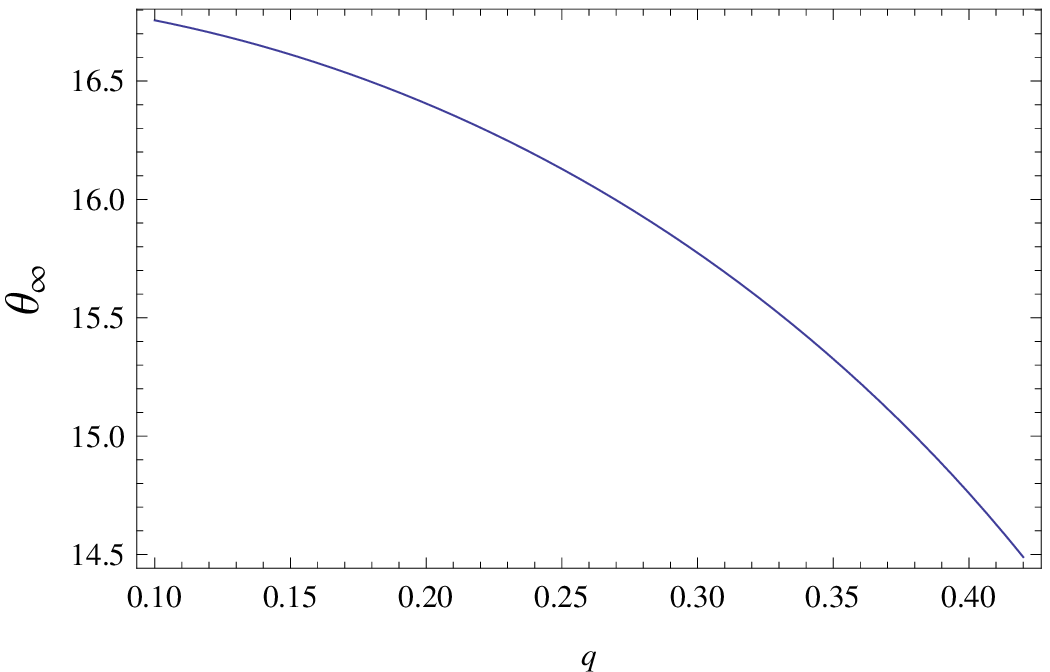}\;\;\;\;\;
 \includegraphics[width=5.0cm]{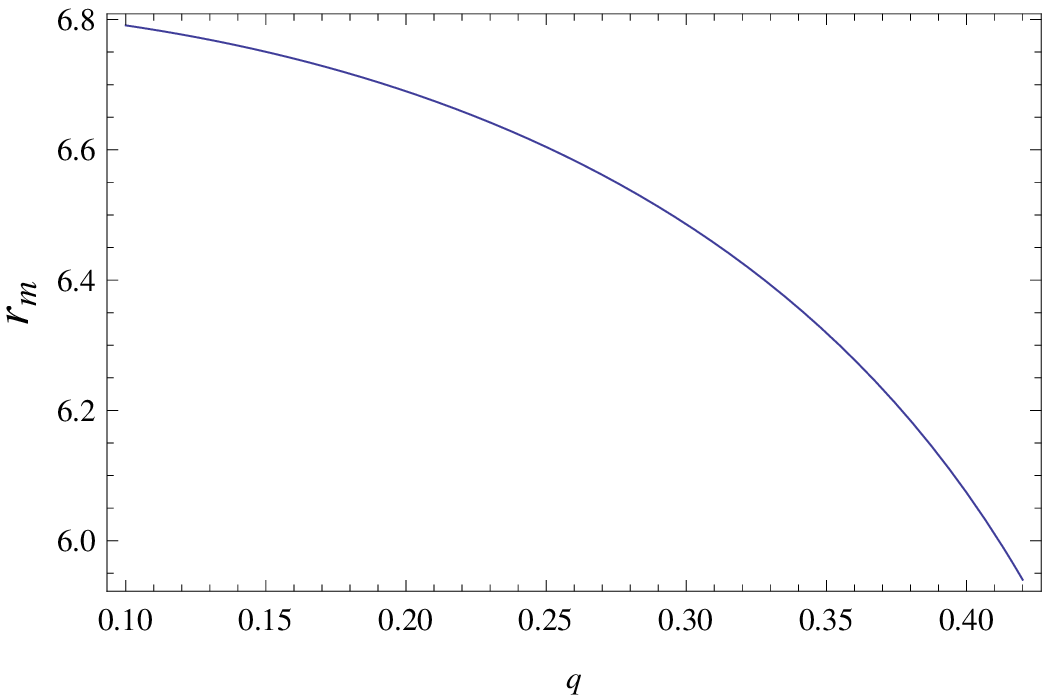}\;\;\;\;\;
 \includegraphics[width=5.0cm]{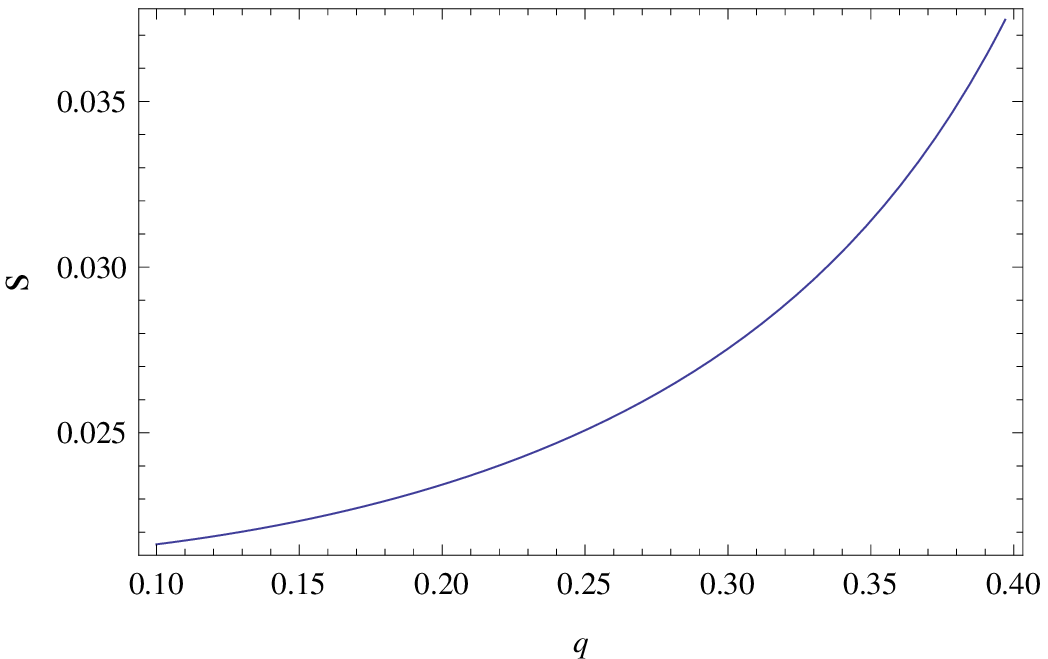}
\caption{Strong gravitational lensing by the Galactic center black
hole. Variation of the values of the angular position
$\theta_{\infty}$, the relative magnitudes $r_m$ and the angular
separation $s$ with parameter $\sqrt{\vartheta}$ in the
noncommutative Schwarzschild black hole spacetime (in the upper row)
and with $q$ in the Reissner-Norstr\"{om} black hole (in the lower
row).}
 \end{center}
 \label{f3}
 \end{figure}
It is easy to obtain that our results reduce to those in the
commutative Schwarzschild black hole sacetime as
$\vartheta\rightarrow0$. Moreover, from the table (II), we also find
that as the parameter $\vartheta$ increases, the minimum impact
parameter $u_{ps}$, the angular position of the relativistic images
$\theta_{\infty}$ and the relative magnitudes $r_m$ decrease, but
the angular separation $s$ increases.

From Fig. (4), we find that in the  noncommutative Schwarzschild
black hole with the increase of parameter $\vartheta$, the angular
position $\theta_{\infty}$ and magnitudes $r_m$ decreases more
slowly, angular separation $s$ increases more slowly than those in
the Reissner-Norstr\"{om} black hole spacetime. This means that the
bending angle is smaller and the relative magnification of the
outermost relativistic image with the other relativistic images is
bigger in the noncommutative Schwarzschild black hole spacetime. In
order to identify the nature of these two compact objects lensing,
it is necessary for us to measure angular separation $s$ and the
relative magnification $r_m$ in the astronomical observations.
Tables (II) tell us that the resolution of the extremely faint image
is $\sim 0.03$ $\mu$ arc sec, which is too small. However, with the
development of technology, the effects of the spacetime
noncommutative constant $\vartheta$ on gravitational lensing may be
detected in the future.

\section{Summary}

Noncommutative geometry may be a starting point to a quantum gravity.
Spacetime noncommutative constant would be a new fundamental natural constant
which can affect the classical gravitational effect such as gravitational lensing.
 Studying the strong gravitational lensing can help
us to probe the spacetime noncommutative constant and the
noncommutative gravity. In this paper we have investigated strong
field lensing in the noncommutative Schwarzschild black hole
spacetime to study the influence of the spacetime noncommutative
parameter on the strong gravitational lensing. The model was applied
to the supermassive black hole in the Galactic center. Our results
show that with the increase of the parameter $\vartheta$ the minimum
impact parameter $u_{ps}$, the angular position of the relativistic
images $\theta_{\infty}$ and the relative magnitudes $r_m$ decrease,
and the angular separation $s$ increases. Comparing to the
Reissner-Norstr\"{om} black hole, we find that the angular position
$\theta_{\infty}$ and magnitude $r_m$ decrease more slowly, angular
separation $s$ increases more slowly. In a word, the influences of
spacetime noncommutative parameter are similar to those of the
charge, just they are much smaller. This may offer a way to
distinguish a noncommutative Schwarzschild black hole from a
Reissner-Norstr\"{om} black hole by the astronomical instruments in
the future.

\begin{acknowledgments}
This work was partially supported by the Scientific Research
Foundation for the introduced talents of Hunan Institute of
Humanities Science and Technology. S. Kang's work was supported by
the National Natural Science Foundation of China (NNSFC)
No.10947101; C.-Y. Chen's work was supported by the NNSFC
No.11074070; J. Jing's work was supported by the NNSFC No.10675045,
No.10875040 and No.10935013, 973 Program No. 2010CB833004 and the
HPNSFC No.08JJ3010£» S. Chen's work was supported by the NNSFC
No.10875041, the PCSIRT No. IRT0964 and the construct program of key
disciplines in Hunan Province.
\end{acknowledgments}


\begin{thebibliography}{99}
\bibitem{foot} The notation $\vartheta$ used here is a constant as well as Plank
constant $\hbar$, but we still call it as a spacetime noncommutative
parameter since it up to now is undetermined.
\end{thebibliography}
\end{document}